# Practical method to reclassify Web of Science articles into unique subject categories and broad disciplines

Staša Milojević

Center for Complex Networks and Systems Research, Luddy School of Informatics, Computing, and Engineering, Indiana University, Bloomington

Email: smilojev@indiana.edu

## Abstract

**Classification of bibliographic items into subjects and disciplines in large databases is essential for many quantitative science studies. The Web of Science classification of journals into ~250 subject categories, which has served as a basis for many studies, is known to have some fundamental problems and several practical limitations that may affect the results from such studies. Here we present an easily reproducible method to perform reclassification of the Web of Science into existing subject categories and into 14 broad areas. Our reclassification is at a level of articles, so it preserves disciplinary differences that may exist among individual articles published in the same journal. Reclassification also eliminates ambiguous (multiple) categories that are found for 50% of items, and assigns a discipline/field category to all articles that come from broad-coverage journals such as *Nature* and *Science*. The correctness of the assigned subject categories is evaluated manually and is found to be ~95%.**

## 1. Introduction

The problem of the classification of science has entertained the attention of philosophers and scientists alike for centuries (Dolby, 1979). The practice of classification is usually understood as a process of arranging things "in groups which are distinct from each other, and are separated by clearly determined lines of demarcation" (Durkheim & Mauss, 1963, p. 4). However, nature, and therefore science, with all its complexity, does not conform to any particular categorization or hierarchical structuring (Bryant, 2000) and there is no singular or perfect classification (Glänzel & Schubert, 2003). Despite inherent limitations, classifications are of practical use to organize and study knowledge. Many classification schemes of science and scientific literature have been proposed, with different levels of granularity and/or hierarchy. Different schemes have different levels of complexity and sophistication, and criteria can be constructed to compare and evaluate them (Rafols & Leydesdorff, 2009).

The classification of scientific literature has been pursued within quantitative science studies since at least the 1970s (e.g., Carpenter & Narin, 1973; Narin, Carpenter, & Berlt, 1972; Small & Griffith, 1974; Small & Koenig, 1977). A number of studies frame this research as discipline/field delineation or delimitation (Gläser, Glänzel, & Scharnhorst, 2017; Gómez, Bordons, Fernandez, & Méndez, 1996; López-Illescas, Noyons, Visser, De Moya-Anegón, & Moed, 2009; Zitt, 2015). The search for adequate solutions to classification has intensified in recent years, often motivated by finding appropriate reference sets for citation normalization needed for evaluation studies (Bornmann, 2014; Glänzel & Schubert, 2003; Haunschild, Schier, Marx, & Bornmann, 2018; Leydesdorff & Bornmann, 2016).

Recent classification efforts have most commonly been divided into journal-focused and paper (article)-focused solutions. The most prevalent and widely used classification of literature into disciplines is via journals, based on a simplistic assumption that a discipline can be defined through journal subject categories (Carpenter & Narin, 1973; Narin, 1976; Narin, Pinski, & Gee, 1976). Such approach is not surprising – journals often serve as anchors for individual research communities, and new journals may signify the formations of disciplines. On a more practical note, the Web of Science (WoS) Journal Citation Reports subject categories is "one of the few classification systems available, spanning all disciplines" (Rinia, van Leeuwen, Bruins, van Vuren, & Van Raan, 2001, p. 296) and is easy to implement since it is available for items in one of the most widely used bibliographic databases, WoS. WoS classifies all of the journals it indexes into ~250 groups called *subject categories*. Each journal is classified into one, or up to six, subject categories. The classification uses a number of heuristics and its rather general description is provided by Pudovkin and Garfield (2002). WoS classification is not explicitly hierarchical, even though some subject categories can be considered as part of other, broader ones. In addition, WoS contains categories that are explicitly broad (labeled as 'multidisciplinary') in order to describe the content of journals that publish across one broad area or across entire science.

Over the years, a number of other journal-centered classifications have been developed. Most of them are hierarchical. For example Scopus, another major bibliographic database, uses All Science Journal Classification (ASJC). National Science Foundation (NSF) uses a two-level system in which journals are classified into 14 broad fields and 144 lower-level fields known as CHI, after Narin and Carpenter's company, Computer Horizons, Inc., which developed it in the 1970s (Archambault, Beauchesne, & Caruso, 2011). Science-Metrix uses 3-level classification which classifies journals into exclusive categories using both algorithmic methods and expert judgment (Archambault et al., 2011). Glänzel and Schubert (2003) developed KU Leuven ECOOM journal classification. Gómez-Núñez, Vargas-Quesada, de Moya-Anegón, & Glänzel (2011) used reference analysis to reclassify the SCImago Journal and Country Ranks (SJR) journals into 27 areas and 308 subject categories. Some classifications used a hybrid method combining text and citations to cluster journals (Janssens, Zhang, De Moor, & Glänzel, 2009). Chen (2008) has used WoS as a starting point for developing a classification using affinity propagation method on journal-to-journal citation network. University of California San Diego (UCSD) classification has been developed in mapping of science efforts (Börner et al., 2012).

Journal-level classification suffers from a number of problems, many of which have been pointed out previously. For example, Klavans & Boyack (2017) found journal-based taxonomies of science to be more inaccurate than the topic-based ones and therefore argued against their use. Similar findings were reported in a recent study which carried out direct comparison of journal- and article-level classifications (Shu et al., 2019) reporting that journal-level classifications have the potential to misclassify almost half of the papers. The issues with the accuracy might be tied to the increase both in the number of journals that publish papers from multiple research areas and the number of papers published in those journals, making journal-level classifications problematic (Gómez et al., 1996; Wang & Waltman, 2016). While journal-level classifications underperform compared to article-level classification in micro-level analyses, they might still be useful for (non-evaluative) macro-level analysis (Leydesdorff & Rafols, 2009; Rafols & Leydesdorff, 2009).

The usage of journals as an appropriate level for classification has been problematized even for journals with unique, non-multidisciplinary classification in WoS, given that a journal may publish articles from different disciplines and would not be the right unit to capture interdisciplinary activities (Abramo, D'Angelo, & Zhang, 2018; Klavans & Boyack, 2010). Boyack and Klavans (2011) suggest that "few journals are truly disciplinary" (p. 123). In their study of research specialties, Small and Griffith (1974) found journals to be too broad a unit of analysis and called for the usage of publications instead. The mounting body of research pointing to the drawbacks of journal-based classifications have prompted the development of article-level classifications. These efforts are usually accompanied by the development of new classification schemes, and are often called algorithmic classifications, due to the clustering techniques used to come up with classes/categories (Ding, Ahlgren, Yang, & Yue, 2018). Klavans and Boyack (2010) have pioneered these classifications at large scale using co-citation techniques (bibliographic coupling of references and keywords) at the paper level to develop the SciTech Strategies (STS) schema consisting of 554 topics, and an alternative method based on co-citation analysis of highly cited references to identify over 84,000 paradigms. Further advances in these techniques were made by Waltman and van Eck (2012), who used direct citations with the minimum number of publications per cluster and a resolution parameter to come up with a 3-level classification. Their work has been further advanced by creating a number of algorithmic classifications at different levels of granularity (Ruiz-Castillo & Waltman, 2015) and searches for the optimal resolution parameter for the level of topics (Sjögårde & Ahlgren, 2018). In addition, since these methods are based on clustering algorithms, and it has long been argued that the resulting classifications are not algorithm-neutral (Leydesdorff, 1987), some studies addressed how different algorithms affect resulting classifications (Šubelj, van Eck, & Waltman, 2016). Overall, the article-based classifications have been praised for being able to classify papers regardless of the type of journals they were published in and placing each publication into a single class/category. One of the drawbacks of the paper-level classification is the problem of naming the classes/categories (Perianes-Rodriguez & Ruiz-Castillo, 2017) making these classifications problematic for macro-level analysis (Ding et al., 2018).

The usefulness of classification schemes for science studies and research evaluation is not determined only by its quality, by also by the availability of a classification of scientific literature at all levels of analysis (from micro to macro), flexibility for different purposes, and the simplicity of interpretation and reproduction. While it is clear that journal-level classifications in general, and WoS journal-level classification in particular, have a number of shortcomings, they are still widely used, primarily because of their wide availability and the familiarity of audiences with WoS subject categories. An article-level classification that would still use the familiar WoS subject categories would be a welcome and practical solution to some of the problems of journal-level classification, but no such classification currently exists. The purpose of this work is to fill this gap by presenting a flexible, simple and easily reproducible method to reclassify WoS items using existing WoS categories, but at the article level. Such a classification is particularly useful for "descriptive bibliometrics" (Borgman & Furner, 2002) or "science of science" (Fortunato et al., 2018) research, especially when the comparison across all the fields and over long time periods is needed.

In addition to being journal-level there are two additional practical problems with WoS classification that will be addressed in the proposed reclassification. One problem is related to different levels of specialization of journals (Glänzel, Schubert, & Czerwon, 1999). The scope of journals ranges from highly specialized ones, via those that cover a whole range of subfields within a field or a discipline (e.g., general journals in physics, chemistry etc.), to journals covering multiple disciplines or fields (Narin, 1976). In WoS subject categories, journals that cover entire large disciplines (broader than typical subject categories) are classified as "multidisciplinary" (e.g., "Physics, multidisciplinary" includes journals containing individual articles actually belong to specific subject categories, such as "Physics, nuclear"; "Optics"; "Thermodynamics", etc.). In addition, there are journals such as *Nature*, *Science* and *PNAS* which cover many disciplines, and are classified in WoS as "Multidisciplinary Sciences". Such journals rarely carry truly muldisciplinary articles, but rather articles from a large number of disciplines (Katz & Hicks, 1995; Waltman & van Eck, 2012). Altogether, 10% of WoS items belong to nine explicit "multidisciplinary" categories. Without the means to establish their true subject category, these articles are often excluded from the analyses of disciplinary practices, thus removing what are often the articles with very high impact (Fang, 2015). As a solution to this problem a number of researchers suggested reclassification of individual articles in such journals, especially in the subject category "Multidisciplinary Sciences". Many of the proposed solutions are based on the references of the articles (e.g., Glänzel & Schubert, 2003; Glänzel, Schubert, & Czerwon, 1999; Glänzel, Schubert, Schoepflin, & Czerwon, 1999; López-Illescas et al., 2009). A more recent solution to this problem utilized both citing and cited publications as a basis for reclassification (Ding et al., 2018). Our article-level reclassification of WoS classifies articles from such multidisciplinary journals into other more specific WoS subject categories.

The second problem of WoS classification is the lack of exclusivity (Bornmann, 2014; Herranz & Ruiz-Castillo, 2012a, 2012b). Namely, many journals in WoS (containing, by our estimate, 40% of all items in WoS) are assigned more than one subject category (in agreement with other studies, e.g., Herranz and Ruiz-Castillo (2012a), who reported that 42% of 3.6 million articles published in 1998-2002 were assigned to more than one category, and Wang and Waltman (2016) who reported that almost 60 percent of journals in WoS are assigned a single category). Multiple subject categories lead to ambiguities when it comes to the analysis. Should such articles be counted in each category, artificially increasing their weight in the overall analysis? Should they be counted fractionally, and thus decreasing their weight within a single category? How to treat them when a non-overlapping delineation is desired, as is often the case? Most journals are assigned multiple categories because they cover more than one subject, even though articles in them usually deal predominantly with one subject. Less often the articles, and not just the journal, are indeed positioned at the intersection of several subjects, and multiple subjects may be appropriate. In such cases we may still wish to assign a primary single category to arrive at non-overlapping delineation of scientific literature. As in the case of "multidisciplinary" categories, references have been proposed for the classification of journals (and articles) with multiple WoS categories into unique categories (e.g., Glänzel & Schubert, 2003; Glänzel, Schubert, & Czerwon, 1999; Narin, 1976; Narin et al., 1976). Our article-level reclassification will assign the most prevalent subject category as the single category for each article and remove the ambiguity. Information regarding potential multidisciplinarity at the level of article will nevertheless be retained if required for the analysis.

Finally, many of the large-scale studies, especially the ones that are comparative in nature, require a smaller number of broader classes. To achieve this goal we additionally categorize articles into 14 broad areas, based on NSF WebCASPAR classification (Javitz et al., 2010).

## 2. Proposed approach

In this paper we propose a *reference*-based (re)classification system that can easily be applied at various levels of granularity. The approach is relatively straightforward and allows for easy reproducibility. Also, by using existing WoS subject categories as units of classification the approach obviates the need to develop an independent scheme for defining and naming of the classes/categories.

Following previous efforts, our approach is to use each item's references to infer the topic of a bibliographic item. However, given the problems identified above, we initially use only references that were published in journals that have a single subject category which is not "multidisciplinary" (i.e., it is not published in multidisciplinary or general disciplinary journals). Such an approach appears appropriate given that previous studies have found WoS subject categories to be fairly precise description of subjects of individual articles published in journals described with one or two subject categories (Glänzel, Schubert, & Czerwon, 1999; Glänzel, Schubert, Schoepflin, et al., 1999) and that central journals within particular disciplines "exhibit little cross citing" (Narin, 1976, p. 194).] For the purposes of this paper, we refer to such items as *classifier references* or *classifiers*. The tallying of the subject categories of classifier references allows us to determine the unique WoS subject category of items that originally had multiple categories or were placed in multidisciplinary categories. However, what is novel in our approach is that the method is applied to reclassify all items that contain classifier references, whether they had unique original (journal-based) classification or not, in order to obtain a consistent comprehensive classification at the level of *individual* items, i.e. articles. Also, unlike a number of other approaches, this one does not apply a particular threshold that an item should meet in order to be classified into a particular category (e.g., Fang, 2015; Glänzel, Schubert, & Czerwon, 1999; Gómez-Núñez et al., 2011; López-Illescas et al., 2009), giving every item a definitive category.

The proposed approach allows both for the classification into exclusive classes (where each article is placed into a single class) and, if needed for particular research questions, a construction of a detailed vector description of disciplinary composition of articles (and consequently, of journals, authors, etc.), which will be described in a future work.

In the remainder of the paper we describe the data, methodology and evaluation of the proposed approach using the WoS. The approach itself is rather general and similar methodology can be used both to reclassify articles in WoS using different starting classification of core journals or classifying articles in other databases that use journal-level classifications. We present the results of the classification of individual items both at the level of subject categories and an aggregated level of broad research areas. New classifications are evaluated using an automated method and validated using blind manual classification.

# 3. Data and methodology

## 3.1. Initial reclassification

For (re)classification we use the full Web of Science (WoS) Core Collection database containing items published from 1900 through the end of 2017. The database contains a total of 69 million items (bibliographic entries), of which 55 million have at least one reference recorded in the database. WoS items belong to different document types such as articles, proceedings papers, editorials, letters, reviews, etc. We perform the classification on (and using) all document types, but carry out the evaluation and validation on document types *article* and *proceedings paper* – the items containing original research and most often used in analyses. There are 45 million items of these two types in WoS with at least one reference, and we refer to them collectively as just the "articles". The edition of WoS used in this work uses 252 subject categories. Classification was extracted from the SQL table `subjects` using the subject category collection referred to by field `ascatype` as the 'traditional'. Categories are listed in Table A in the Appendix.

For higher-level classification, we place each of 252 subject categories into 14 broad areas. Names of broad areas are taken from NSF WebCASPAR Broad Field (Javitz et al., 2010), except that we include their "Other life sciences" within "Medical sciences". Mapping between WoS subject categories and our broad areas, given in Table A, follows Javitz et al. (2010) mapping between ipIQ Fine Field category (formerly CHI category) and WebCASPAR Broad Field whenever there is an ipIQ category that clearly matches WoS category. In other instances (1/2 of all WoS categories) the broad category is determined by the author.

WoS attempts to match each item's references to other items in WoS. It is the items that have matched references that can be reclassified using the proposed method. Furthermore, to allow initial classification using our method, the references need to be *classifiers*, i.e., items whose original classification is unique and non-multidisciplinary. 41 million items contain classifier references and can therefore be classified into subject categories, of which 36 million are articles, representing 79% of all articles with references. We will outline later in this section how this percentage can be further increased using iterative approach. Classification into broader areas is possible for a larger number of items (44 million of any type, and 38 million articles), because classifiers can include items classified as multidisciplinary, as long as they can be placed in some broad area (e.g., category "Physics, multidisciplinary" can be used, but "Multidisciplinary Sciences" cannot.) The fraction of articles (containing references) that can be classified, as a function of publication year, is shown in Figure 1. The fractions are above 90% in recent years, and are relatively high since the 1950s. The rising trend is likely a combination of several factors: more complete efforts on behalf of database administrators to match the references in recent publications, journal articles becoming "the central medium for the dissemination and exchange of scientific ideas" (Bowker, 2005, p. 126), and the overall increase in the number of references per paper over time (Milojević, 2012; Price, 1963; van Raan, 2000), all of which increase the chances of an article containing classifier references. The items that remain without new classification are rarely full-fledged research papers, but most often items such as book reviews or short conference proceedings.

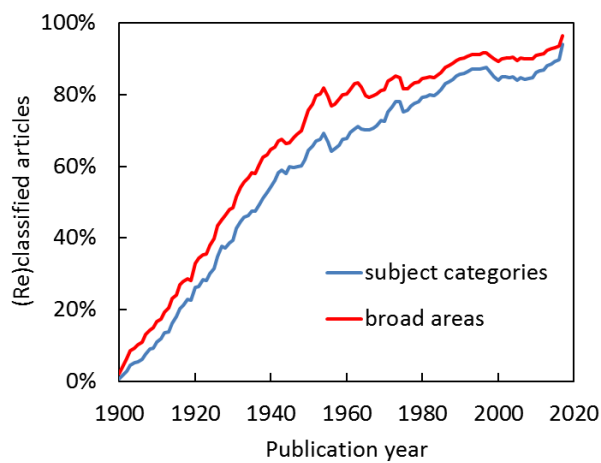

Figure 1. Percentage of all articles containing references that can be reclassified into subject categories or broad areas as a function of article publication year. Numbers are based on initial reclassification. An iterative pass will increase the percentage of articles classified into subject categories by 5%.

For classification at the subject category level, 20 million items serve as classifiers. Algorithm for the entire classification procedure is given in Figure 2. Classification at the level of subject categories proceeds as follows. For each classifiable item we go through all of its classifier references and produce a ranked list of their subject categories. A subject category that is the most frequent is adopted as a new (reclassified) subject category. Most often the distribution of categories is dominated by the most frequent subject category (the article is predominantly unidisciplinary). Occasionally, the tallying results in a tie between two most frequent categories (13% of cases). We attempt to break the ties by adding to the tally the original subject category (or categories, if they were multiple). This can be done if the original subject category is non-multidisciplinary. 52% of the ties can be broken in this way. Otherwise, we adopt as the final classification the category with a larger number of articles.

The granularity of reclassified subject categories defined as the number of items divided by the sum of the items in each category squared (Waltman, Boyack, Colavizza, & Van Eck, 2019) is $1.5 \times 10^{-6}$, compared to $2.3 \times 10^{-6}$ for the original classification, i.e., it is relatively similar. The number of categories of different sizes (i.e., total number of reclassified items) is presented in Figure 2. Categories span a very wide range of sizes.

Classification at the level of broad areas proceeds in the same way, except that the ranked list is made of classifiers' broad areas. For classification into broad areas, the number of classifiers is 50% larger than in the case of subject categories, 30 million, because individual subject categories of items that have multiple subject categories most often belong to the same broad area, and such items are therefore eligible to serve as classifiers. For the classification of items into broad areas ties happen in 4% of all cases, and can be resolved by including the original broad area in the ranked list in 69% of those cases. Otherwise, we take the more populous category as the final one.

Overall, the classification is not very sensitive to the extent of the classifier set. We perform the test in which we base the classification on only 1/2 of all available classifiers. The resulting broad categories agree with the ones obtained with the full classifier set in 94% of cases.

The exact counts pertaining to the dataset and initial reclassification are provided in Tables 1 and 2.

Table 1. Number of items from the Web of Science used in (re)classification.

|  | All types | Articles + Conference Proceedings |
|---|---|---|
| All items | 69,326,147 | 49,775,351 |
| with references | 54,581,163 | 45,219,572 |
| multidisciplinary | 5,585,211 | 4,640,854 |
| multidisc. science | 1,317,033 | 1,071,437 |

Table 2. Number of classified items of different types after initial reclassification. Percentage in parentheses is with respect to all such items with references.

|  | Subject category classification | | Broad area classification | |
|---|---|---|---|---|
|  | All types | Articles + Conference Proceedings | All types | Articles + Conference Proceedings |
| Classifier items | 20,286,801 |  | 29,853,395 |  |
| Classified items | 41,132,197 (75%) | 35,940,588 (79%) | 43,847,374 (80%) | 38,118,382 (84%) |
| multidisciplinary | 3,719,208 (67%) | 2,599,373 (56%) |  |  |
| multidisc. science | 896,169 (68%) | 740,592 (69%) | 909,543 (69%) | 792,875 (74%) |

```
LOAD DATA: All WoS items
        WOS_ID # WoS identifier
        CLASS          # Classification*
*      = Original WoS subject categories (one or more)    OR
       = Original WoS subject categories converted into one or more broad areas

[ Optional, for iterative (second) pass
[LOAD DATA: All reclassified WoS items
[      WOS_ID # WoS identifier
[      NEW_CLASS      # Classification obtained in first pass of this algorithm

LOAD DATA: References of all WoS items
        WOS_ID # Citing WoS identifier
        CWOS_ID# WoS identifier of cited item

[      IF NEW_CLASS(CWOS_ID) exists
[            CLASS(CWOS_ID) = NEW_CLASS(CWOS_ID) # Replace with new classification
       IF CLASS(CWOS_ID) exists AND unique AND NOT multidisciplinary # Classifier reference
            DIST(WOS_ID)+=1 # Increment distribution of categories/areas for a given WoS
item
            COUNT(CLASS)+=1 # Increment counter for this category/area

LOOP: All WoS items with classifier references
       SORT DESCENDING DIST(WOS_ID) # Determine 1st and 2nd  most frequent category/area
(CLASS1/CLASS2)
       IF N(CLASS1) = N(CLASS2)  # Tie
            LOOP: CLASS(WOS_ID) # Original categories for the item
                  DIST(WOS_ID)+=1 # Add them to distribution
            SORT DESCENDING DIST(WOS_ID) # Redetermine most frequent category
                  IF N(CLASS1) = N(CLASS2)  # Still a tie
                        SORT DESCENDNG COUNT(CLASS1,CLASS2) # Take bigger class as the
final class
       NEW_CLASS(WOS_ID) = CLASS1
       OUTPUT NEW_CLASS(WOS_ID)
```

Figure 2. An algorithm (pseudo code) describing the reclassification procedure.

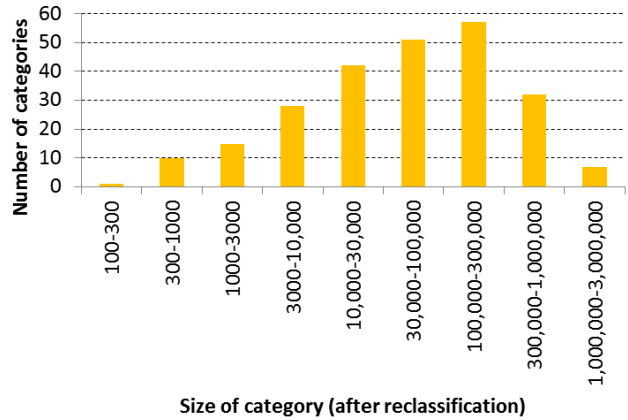

Figure 3. Size distribution of WoS subject categories after initial reclassification.

## 3.2 Iterative reclassification

Once the reclassification has been carried out, it is possible and often recommended to carry out the process of reclassification iteratively. In iterative reclassification the tallying of subject categories of references and the determination of which reference can serve as classifier is based on the reclassified subject categories (or broad areas, for the high-level classification). The process can be repeated multiple times, but here we limit ourselves to one iterative pass and the quality and extensiveness of this second reclassification compared to the first. Iterative pass is procedurally very similar to the original one, and the needed modifications are laid out in Figure 2. After the iterative pass 9% of items acquire a different broad-area classification, and 20% of items acquire a different subject category.

There are two principal reasons for carrying out the iterative pass: an increase in the number of items that can be classified, and, potentially, an increased accuracy of new categories. In the original pass only items that had classifier references could be classified, which, as we have shown, represents 79% of all articles, and around 90% of recent articles. Items that only have had references with multiple original categories and/or multidisciplinary categories could not be classified. However, after the first reclassification, most of these references will receive unique, non-multidisciplinary classification and can now serve as classifiers. The numbers of items and articles that can be classified in the iterative pass is presented in Table 3. Comparing these numbers to those in Table 2 we see a relatively significant increase in the number of items or articles that get classified into subject categories (~8%) and a more modest increase of items/articles classified into broad areas (~2%).

Table 3. Number of classified items of different types after the second (iterative) reclassification.

| | Subject category classification | | Broad area classification | |
|---|---|---|---|---|
| | All types | Articles + Conference Proceedings | All types | Articles + Conference Proceedings |
| Classifier items | 36,104,403 | | 38,504,614 | |
| Classified items | 44,349,678 (81%) | 38,450,585 (85%) | 44,936,331 (80%) | 38,918,386 (84%) |
| multidisciplinary | 4,317,080 (77%) | 2,931,707 (63%) | | |
| multidisc. science | 1,011,770 (77%) | 804,203 (75%) | 968,783 (74%) | 822,849 (77%) |

The increase of completeness using the iterative pass is especially significant in the cases where the majority of the journals in some discipline originally had multiple WoS categories, and were therefore precluded from serving as classifier references. While such cases are not common in general, one of them happens to include core journals in quantitative studies of science. Namely, *Journal of Informetrics* (JoI), *Scientometrics* and *Journal of the Association for Information Science and Technology* (JASIST) are all listed with two WoS subject categories: "Computer Science, Interdisciplinary Applications" and "Information Science & Library Science", which means that they cannot serve as classifiers, at least not in the initial pass. For example, out of 840 items published in JoI, 663 can be classified in the first pass (79%), a lower fraction than on average. Interestingly, of the classified items, 41% received the classification of "Information Science & Library Science", while essentially none were classified as "Computer Science, Interdisciplinary Applications". This shows that the reclassification successfully rejected this obviously inappropriate categorization. In the iterative pass, however, the number of classified articles increased substantially, to 796 (95% of total). Furthermore, 52% has now received the classification of "Information Science & Library Science", most of any category. Other frequent categories included "Economics" (9%), "History and Philosophy of Science" (8%) and "Sociology" (6%).

To conclude, extending the classification to include the iterative pass provides increase in the number of classified items (especially at subject category level), which for certain cases can be quite significant.

## 4. Validation and evaluation

The validation and evaluation of the approach and of the final reclassification is performed using three tests, each serving a separate purpose:

a) Automatic internal test against the original WoS classification, in order to validate the methodology.

b) Manual tests in order to evaluate the accuracy of reclassification in comparison to the original WoS classification.

c) Manual external test in order to evaluate the overall reliability of the resulting classification.

a) To validate the methodology and hone the approach we have performed an automatic test by calculating the percentage of articles whose original and new classifications agree. This test can only be performed on items whose original classification was unique and non-multidisciplinary. This test is internal because we do not evaluate the accuracy of the original WoS classification using any external knowledge.  We do not expect the test to produce 100% agreement. First of all, the reclassification is at the level of articles, whose topics may be to some extent different from those of their journals, and second, because the subject categories are rarely entirely mutually exclusive, so a reclassified category may be related, but not exactly the same as the original one. The value of this test is in the relative assessment. When evaluating, for example, two article-level classification schemes, the one that has a higher level of agreement with respect to, however imperfect, reference classification (in this case the original classification), should be considered more accurate internally. For the reclassification at the level of subject categories we find the overall agreement to be 66% after the initial reclassification and 58% after the iterative pass.  In comparison, an alternative classification scheme that we devised but ultimately did not adopt, which uses the similarity of titles to perform reclassification, had an agreement of <50%. For this alternative method we calculated TF-IDF ("term frequency-inverse document frequency") values between each article title to be reclassified and each of the classifier articles (articles that have unique non-multidisciplinary WoS category).  In this case, IDF actually represents inverse title word frequency that was first determined from the entire dataset, and TF-IDF is the sum of all IDFs of the words that overlap.  For an article to be classified we adopt the category of an article with the greatest TF-IDF value.

The level of agreement varies from one subject category to another. It is the highest for astronomy and astrophysics – 97%. The number of articles in different categories varies widely, with the largest category being 2000 times larger than the smallest (see Figure 3). We find that the agreement is correlated with the size of the subject category, with larger categories having a higher level of agreement. This is probably because some of the smaller categories can also be considered subcategories of larger ones, so many of the articles get reclassified into these larger categories. The opposite (an item that was originally in a larger category being reclassified into a smaller one) is less likely simply because there are fewer classifiers that belong to smaller categories. Furthermore, small categories may represent more recent disciplines, which would naturally cite works from the disciplines from which they emerged. As we will see shortly, this lower level of agreement for smaller categories does not imply that the new category is incorrect – it may simply be placing individual items in a related, equally correct  subject category, or may reflect a high degree of the interdisciplinarity of an article.

We perform a similar automatic validation for broad-area classification and find the overall agreement of 85% after the initial reclassification and 82% after the iterative pass. Agreement in different areas is now more similar, ranging from 60% for agricultural sciences (which tends to be highly interdisciplinary) to 93% for astronomy and astrophysics (which has a very low degree of interdisciplinarity), and the level of agreement is not correlated with the size of the area.

b) The internal validation in itself does not allow us to evaluate the *quality* of the reclassification with respect to the original classification.  We assess this by manual evaluation, performed by the author, in the following way. For 142 randomly selected articles whose original classification was unique and non-multidisciplinary we output the original and new subject category. The order in which the two categories are written out is randomly reversed in 50% of cases. The evaluator does not know a priori which category is original and which is new – this information is saved separately and is used only after the evaluation was performed. Evaluator's task is to select the subject category that better describes the article based on its title (and abstract, if necessary), but ignoring the name of the journal, as not to bias the assessment, since the journal topic was the basis for the original classification. If both categories are estimated to be equally appropriate, this is also indicated. After the initial reclassification 91 out of 142 articles had the same new and old category (64%; in agreement with the full sample). For 25 articles old and new categories were equally good (most often because one category can be considered a part of another). Of the remaining 26 articles the original classification was considered better in 15 cases and the new one in 11 cases. In 15 cases where the original classification was considered better, the new one was still essentially correct in 13 cases. Altogether, the initial reclassification is nearly as good as the original one, i.e., we have not introduced spurious results in the process of reclassification. The differences between original and new classification revealed by automated validation can be attributed to article's interdisciplinarity (such that both categories are correct) and to somewhat stratified, non-exclusive nature of WoS subject categories (again making both categories correct).

Manual evaluation is also carried out for the same 142 articles for their broad-area classifications. Areas agree for 124 articles (87%; in agreement with the full sample), and are considered equally good in 4 cases. Of the remaining 14 articles the original classification is considered better in only 3 cases, while the new area is considered more accurate in the remaining 11 cases, i.e., the new classification is overall somewhat better.

c) The overall reliability of the new classification is what is ultimately of most interest. We test it based on an external assessment, which looks at all items irrespective of how the items were originally classified, i.e., it includes items that originally had ambiguous classification or where the classification was effectively missing because the item was published in a multidisciplinary journal. The test is performed by the author by evaluating the correctness of subject categories and broad areas of 100 randomly selected items, based on their titles and abstracts. We find 92% of subject categories and 95% of broad areas to be correct after the initial reclassification. The accuracy increased to 95% for subject categories and 97% for broad areas after the iterative pass. It needs to be pointed out that whereas the error rate is relatively small across the entire dataset, it need not be uniform in different disciplines or for different journals, so it is advisable to perform similar manual tests for subsets of dataset that one wishes to study.

# 5. Extension of the method using citation data

It is in principle possible to adapt our method to use not only the references as the basis for reclassification, but also the citations. Citations, at least in the initial reclassification, would also have to come from sources that have unique, non-multidisciplinary WoS category. The use of citations may allow some items to be classified that otherwise did not have classifier references. We carry out such reclassification at broad-area level and find that the number of classified items increases from 43,847,374 (63% of all possible items, regardless of whether they had references or not) to 47,593,363 (69%). The increase exceeds that from the iterative pass (44,936,331 or 65%). The fraction is still short of 100% because most of the items that lack references also lack citations (most of them are not really citable items.) One possible drawback of using citations is the disproportionality of information available for different items. Unlike references, the number of which tends to be normally distributed, the citations follow a power law distributions, with most articles having very few citations and very few having thousands. Furthermore, citations constantly change, making the proposed procedure essentially non reproducible.

There are 6% of articles with no linked references or citations. These are mostly items more than half a century old. For these items one could apply the TF-IDF method that we discussed in Section 4, which has 100% completeness.

## 6. Discussion and conclusion

This paper proposes a method of classification that is based on references and applies it to classifying WoS articles, both at the field and broad research area levels. While some of the previously proposed clustering-based methods may lead to a better delineation, especially for citation normalization, the proposed method has a number of advantages: it is easily replicated and utilizes widely used WoS subject categories and NSF broad subject areas, does not require extensive computational resources (~40 million articles can be classified on a personal computer within several hours), and avoids the problem of naming classes/categories (something that article-level classifications have struggled with, but are making progress on due to more sophisticated natural language processing approaches and including a wider range of fields of bibliographic records). The major purpose for this classification is devising a flexible and simple way of classifying all of the WoS literature for the purposes of "descriptive bibliometrics" or "science of science" studies. The classification has not been designed for the purposes of research evaluation and if used in that context may be outperformed by approaches that identify more focused comparison sets, as in, e.g., Colliander & Ahlgren (2019).

The major limitations of the proposed method are tied to its usage of WoS subject categories as a starting point and references as a major source of data. Since it uses WoS subject categories as seeds, the proposed classification will inherit some of the known problems of this classification, primarily having to do with erroneous lumping of unconnected journals into a single category. This limitation can potentially be alleviated by the iterative procedure. Furthermore, since the method is based on references, it can be applied only to the items that have references. This should not be a problem with most contemporary original research, but may prove a bit problematic for other types of contributions and for older items. At the same time, relying on references, rather than the citations as in some other studies has some advantages, since more articles have cited other works than being cited themselves. This should lead to a higher recall than citation-based classifications have. An approach that combines references and citations is also possible and was described.

Overall, we find the error rate of the resulting classification to be relatively low (<5%) making it a reasonably reliable basis for a wide range of studies. However, the accuracy may be higher or lower for specific research areas, so like with any classification, the users should exercise caution and validate the classification for the sample of interest. Also, as we have pointed out, especially at the level of 252 subject categories, it is often the case that more than one category is essentially correct, so it is advisable to consider all potentially relevant categories when the recall of a sample is important. This is less of an issue for broad areas.

## Acknowledgement

This work uses Web of Science data by Clarivate Analytics provided by the Indiana University Network Science Institute and the Cyberinfrastructure for Network Science Center at Indiana University.

## Author contributions

Staša Milojević: : conceptualization, data curation, formal analysis, methodology, writing

## Competing interests statement

No competing interests to declare.

## Data availability statement

The data used in this paper is proprietary and cannot be posted in a repository.

## Funding information

This material is partially based upon work supported by the Air Force Office of Scientific Research under award number FA9550-19-1-0391.

# Appendix

Table A. The list of WoS subject categories and corresponding broad areas.

| WoS Subject Category | Broad Area |
|---|---|
| Agriculture, Dairy & Animal Science | Agricultural sciences |
| Agriculture, Multidisciplinary | Agricultural sciences |
| Agronomy | Agricultural sciences |
| Fisheries | Agricultural sciences |
| Food Science & Technology | Agricultural sciences |
| Forestry | Agricultural sciences |
| Green & Sustainable Science & Technology | Agricultural sciences |
| Horticulture | Agricultural sciences |
| Astronomy & Astrophysics | Astronomy |
| Anatomy & Morphology | Biological sciences |
| Biochemical Research Methods | Biological sciences |
| Biochemistry & Molecular Biology | Biological sciences |
| Biodiversity Conservation | Biological sciences |
| Biology | Biological sciences |
| Biophysics | Biological sciences |
| Biotechnology & Applied Microbiology | Biological sciences |
| Cell & Tissue Engineering | Biological sciences |
| Cell Biology | Biological sciences |
| Developmental Biology | Biological sciences |
| Ecology | Biological sciences |
| Entomology | Biological sciences |
| Evolutionary Biology | Biological sciences |
| Genetics & Heredity | Biological sciences |
| Microbiology | Biological sciences |
| Mycology | Biological sciences |
| Nutrition & Dietetics | Biological sciences |
| Ornithology | Biological sciences |
| Paleontology | Biological sciences |
| Parasitology | Biological sciences |
| Physiology | Biological sciences |
| Plant Sciences | Biological sciences |
| Reproductive Biology | Biological sciences |
| Virology | Biological sciences |
| Zoology | Biological sciences |
| Chemistry, Analytical | Chemistry |
| Chemistry, Applied | Chemistry |
| Chemistry, Inorganic & Nuclear | Chemistry |
| Chemistry, Medicinal | Chemistry |
| Chemistry, Multidisciplinary | Chemistry |
| Chemistry, Organic | Chemistry |
| Chemistry, Physical | Chemistry |
| Crystallography | Chemistry |
| Electrochemistry | Chemistry |
| Polymer Science | Chemistry |
| Spectroscopy | Chemistry |
| Computer Science, Artificial Intelligence | Computer sciences |
| Computer Science, Cybernetics | Computer sciences |
| Computer Science, Hardware & Architecture | Computer sciences |
| Computer Science, Information Systems | Computer sciences |
| Computer Science, Interdisciplinary Applications | Computer sciences |
| Computer Science, Software Engineering | Computer sciences |
| Computer Science, Theory & Methods | Computer sciences |
| Medical Informatics | Computer sciences |
| Agricultural Engineering | Engineering |
| Automation & Control Systems | Engineering |
| Construction & Building Technology | Engineering |
| Energy & Fuels | Engineering |
| Engineering, Aerospace | Engineering |
| Engineering, Biomedical | Engineering |
| Engineering, Chemical | Engineering |
| Engineering, Civil | Engineering |
| Engineering, Electrical & Electronic | Engineering |
| Engineering, Environmental | Engineering |
| Engineering, Geological | Engineering |
| Engineering, Industrial | Engineering |
| Engineering, Manufacturing | Engineering |
| Engineering, Marine | Engineering |
| Engineering, Mechanical | Engineering |
| Engineering, Multidisciplinary | Engineering |
| Engineering, Ocean | Engineering |
| Engineering, Petroleum | Engineering |
| Imaging Science & Photographic Technology | Engineering |
| Instruments & Instrumentation | Engineering |
| Materials Science, Biomaterials | Engineering |
| Materials Science, Ceramics | Engineering |
| Materials Science, Characterization & Testing | Engineering |
| Materials Science, Coatings & Films | Engineering |
| Materials Science, Composites | Engineering |
| Materials Science, Multidisciplinary | Engineering |
| Materials Science, Paper & Wood | Engineering |
| Materials Science, Textiles | Engineering |
| Mathematical & Computational Biology | Engineering |
| Medical Laboratory Technology | Engineering |
| Metallurgy & Metallurgical Engineering | Engineering |
| Mining & Mineral Processing | Engineering |
| Nanoscience & Nanotechnology | Engineering |
| Neuroimaging | Engineering |
| Nuclear Science & Technology | Engineering |
| Operations Research & Management Science | Engineering |
| Remote Sensing | Engineering |
| Robotics | Engineering |
| Telecommunications | Engineering |
| Transportation | Engineering |
| Transportation Science & Technology | Engineering |
| Environmental Sciences | Geosciences |
| Environmental Studies | Geosciences |
| Geochemistry & Geophysics | Geosciences |
| Geography, Physical | Geosciences |
| Geology | Geosciences |
| Geosciences, Multidisciplinary | Geosciences |
| Limnology | Geosciences |
| Marine & Freshwater Biology | Geosciences |
| Meteorology & Atmospheric Sciences | Geosciences |
| Mineralogy | Geosciences |
| Oceanography | Geosciences |
| Soil Science | Geosciences |
| Water Resources | Geosciences |
| Archaeology | Humanities |
| Architecture | Humanities |
| Art | Humanities |
| Asian Studies | Humanities |
| Classics | Humanities |
| Cultural Studies | Humanities |
| Dance | Humanities |
| Ethics | Humanities |
| Ethnic Studies | Humanities |
| Film, Radio, Television | Humanities |
| Folklore | Humanities |
| History | Humanities |
| History & Philosophy Of Science | Humanities |
| History Of Social Sciences | Humanities |
| Humanities, Multidisciplinary | Humanities |
| Language & Linguistics | Humanities |
| Literary Reviews | Humanities |
| Literary Theory & Criticism | Humanities |
| Literature | Humanities |
| Literature, African, Australian, Canadian | Humanities |
| Literature, American | Humanities |
| Literature, British Isles | Humanities |
| Literature, German, Dutch, Scandinavian | Humanities |
| Literature, Romance | Humanities |
| Literature, Slavic | Humanities |
| Logic | Humanities |
| Medical Ethics | Humanities |
| Medieval & Renaissance Studies | Humanities |
| Music | Humanities |
| Philosophy | Humanities |
| Poetry | Humanities |
| Religion | Humanities |
| Theater | Humanities |
| Women's Studies | Humanities |
| Mathematics | Mathematical sciences |
| Mathematics, Applied | Mathematical sciences |
| Mathematics, Interdisciplinary Applications | Mathematical sciences |
| Statistics & Probability | Mathematical sciences |
| Allergy | Medical sciences |
| Andrology | Medical sciences |
| Anesthesiology | Medical sciences |
| Audiology & Speech-Language Pathology | Medical sciences |
| Cardiac & Cardiovascular Systems | Medical sciences |
| Clinical Neurology | Medical sciences |
| Critical Care Medicine | Medical sciences |
| Dentistry, Oral Surgery & Medicine | Medical sciences |
| Dermatology | Medical sciences |
| Emergency Medicine | Medical sciences |
| Endocrinology & Metabolism | Medical sciences |
| Gastroenterology & Hepatology | Medical sciences |
| Geriatrics & Gerontology | Medical sciences |
| Health Policy & Services | Medical sciences |
| Hematology | Medical sciences |
| Immunology | Medical sciences |
| Infectious Diseases | Medical sciences |
| Integrative & Complementary Medicine | Medical sciences |
| Medicine, General & Internal | Medical sciences |
| Medicine, Research & Experimental | Medical sciences |
| Microscopy | Medical sciences |
| Neurosciences | Medical sciences |
| Nursing | Medical sciences |
| Obstetrics & Gynecology | Medical sciences |
| Oncology | Medical sciences |
| Ophthalmology | Medical sciences |
| Orthopedics | Medical sciences |
| Otorhinolaryngology | Medical sciences |
| Pathology | Medical sciences |
| Pediatrics | Medical sciences |
| Peripheral Vascular Disease | Medical sciences |
| Pharmacology & Pharmacy | Medical sciences |
| Psychiatry | Medical sciences |
| Public, Environmental & Occupational Health | Medical sciences |
| Radiology, Nuclear Medicine & Medical Imaging | Medical sciences |
| Rehabilitation | Medical sciences |
| Respiratory System | Medical sciences |
| Rheumatology | Medical sciences |
| Sport Sciences | Medical sciences |
| Substance Abuse | Medical sciences |
| Surgery | Medical sciences |
| Toxicology | Medical sciences |
| Transplantation | Medical sciences |
| Tropical Medicine | Medical sciences |
| Urology & Nephrology | Medical sciences |
| Veterinary Sciences | Medical sciences |
| Acoustics | Physics |
| Mechanics | Physics |
| Optics | Physics |
| Physics, Applied | Physics |
| Physics, Atomic, Molecular & Chemical | Physics |
| Physics, Condensed Matter | Physics |
| Physics, Fluids & Plasmas | Physics |
| Physics, Mathematical | Physics |
| Physics, Multidisciplinary | Physics |
| Physics, Nuclear | Physics |
| Physics, Particles & Fields | Physics |
| Thermodynamics | Physics |
| Business | Professional fields |
| Business, Finance | Professional fields |
| Communication | Professional fields |
| Education & Educational Research | Professional fields |
| Education, Scientific Disciplines | Professional fields |
| Education, Special | Professional fields |
| Ergonomics | Professional fields |
| Family Studies | Professional fields |
| Health Care Sciences & Services | Professional fields |
| Hospitality, Leisure, Sport & Tourism | Professional fields |
| Industrial Relations & Labor | Professional fields |
| Information Science & Library Science | Professional fields |
| Law | Professional fields |
| Management | Professional fields |
| Medicine, Legal | Professional fields |
| Primary Health Care | Professional fields |
| Social Work | Professional fields |
| Behavioral Sciences | Psychology |
| Psychology | Psychology |
| Psychology, Applied | Psychology |
| Psychology, Biological | Psychology |
| Psychology, Clinical | Psychology |
| Psychology, Developmental | Psychology |
| Psychology, Educational | Psychology |
| Psychology, Experimental | Psychology |
| Psychology, Mathematical | Psychology |
| Psychology, Multidisciplinary | Psychology |
| Psychology, Psychoanalysis | Psychology |
| Psychology, Social | Psychology |
| Agricultural Economics & Policy | Social sciences |
| Anthropology | Social sciences |
| Area Studies | Social sciences |
| Criminology & Penology | Social sciences |
| Demography | Social sciences |
| Economics | Social sciences |
| Geography | Social sciences |
| Gerontology | Social sciences |
| International Relations | Social sciences |
| Linguistics | Social sciences |
| Planning & Development | Social sciences |
| Political Science | Social sciences |
| Public Administration | Social sciences |
| Social Issues | Social sciences |
| Social Sciences, Biomedical | Social sciences |
| Social Sciences, Interdisciplinary | Social sciences |
| Social Sciences, Mathematical Methods | Social sciences |
| Sociology | Social sciences |
| Urban Studies | Social sciences |